\documentclass[
aps,%
11pt,%
final,
notitlepage,%
oneside,%
onecolumn,%
nobibnotes,%
nofootinbib,%
preprintnumbers,%
superscriptaddress,%
showpacs,%
centertags,%
showkeys,%
amsmath,%
amssymb]{revtex4}

\usepackage{graphicx} 
\usepackage{dcolumn} 

\newcommand{\VV}{{\widetilde V}}

\begin{document}
\strut\hfill PSI-PR-08-04 \\
\title{Renormalization of tensor self-energy in Resonance Chiral Theory\footnote{Presented by J.T. at the Hadron Structure '07, 3-7 September 2007, Modra, Slovakia}}

\author{Karol Kampf}
\affiliation{Institute of
Particle and Nuclear Physics, Charles University, V Hole\v{s}ovi\v{c}k\'ach 2,
CZ-18000 Prague, Czech Republic} \affiliation{Paul Scherrer Institut, W\"urenlingen und Villigen,
Ch-5232 Villigen PSI, Switzerland}

\author{Ji\v{r}i Novotn\'y}
\affiliation{Institute of
Particle and Nuclear Physics, Charles University, V Hole\v{s}ovi\v{c}k\'ach 2,
CZ-18000 Prague, Czech Republic}

\author{Jaroslav Trnka}
\affiliation{Institute of
Particle and Nuclear Physics, Charles University, V Hole\v{s}ovi\v{c}k\'ach 2,
CZ-18000 Prague, Czech Republic}
\date{\today}

\begin{abstract}
We study the problems related to the renormalization of
propagators in Resonance Chiral Theory, concentrating on the case
of vector $1^{--}$ resonances in the antisymmetric tensor
formalism. We have found that renormalization of the divergences
of the self-energy graphs needs new type of kinetic counterterms
with two  derivatives which are not present in the original
leading order Lagrangian. The general form of the propagator for
antisymmetric tensor fields could then contain not only  poles
corresponding to the original
 $1^{--}$ resonance states but also to the additional states with opposite
parity which decouple in the free field limit. In some cases,
these dynamically generated additional states  might be negative
norm ghosts or tachyons.
\end{abstract}



\maketitle

\section{Introduction}

The use of effective field theories for the description of the dynamics of
hadrons has made considerable progress in recent years. In the
low energy region $E<\Lambda_H=1\,{\rm{GeV}}$, the dynamics of the
lowest lying states (the pseudoscalar mesons)  is effectively
described by Chiral Perturbation Theory ($\chi$PT) \cite{Wein,
Gasser1, Gasser2, Op6} based on the spontaneous symmetry breaking
of chiral symmetry of QCD.

In the intermediate energy region $(1\,{\rm GeV}\leq E\leq 2\,{\rm
GeV})$ one uses  the Resonance Chiral Theory (R$\chi$T)
\cite{Ecker1, Ecker2, Knecht, Spain1, Spain2, KNT, KNT2}. R$\chi$T
is based on large-$N_C$ QCD  which partially shares a lot of
interesting properties with the physical $N_C=3$ case. In the leading order of
the $1/N_C$ expansion, the QCD spectrum contains infinite towers
of meson resonances with residual interaction suppressed by powers
of $1/\sqrt{N_C}$. Their dynamics  at the leading order in $1/N_C$
can in principle be described in terms of tree level diagrams
within an effective theory with an infinite number of fields. Such a
theory is not known from first principles; however, it can be
basically  constructed on symmetry grounds and its free parameters
can be fixed by phenomenology. The flavour group of large-$N_C$
QCD is $U(N_f)_L\times U(N_f)_R$ (because of the absence of the axial
anomaly in the large-$N_C$ limit) that is spontaneously broken to
$U(N_f)_V$.

R$\chi$T is the approximation of the large-$N_C$ QCD when only
a finite number of resonances in each channel is included. This
approach is well-established, because for example the ${\cal O}(p^4)$ coupling constants of the $\chi$PT
Lagrangian \cite{Ecker1} are successfully predicted. In addition, there are some developments in saturation of the ${\cal
O}(p^6)$ coupling constants \cite{Karel} with a lot of
phenomenological consequences, see {\em{e.g.}} \cite{Spain1,
Spain2, Roig}.

In this paper we want to show that R$\chi$T might
contain some problems and features of inner inconsistency when we
go behind leading order (quantum loops in R$\chi$T were already
studied in \cite{Loop, Cil}). The more detailed treatise of  the
discussed problem will be published in \cite{KNT3}.

In the following we are interested only in the sector of vector
resonances $1^{--}$ but the results of more general discussion do
not differ from this special case.

\section{Antisymmetric tensor formulation of R$\chi$T}

The standard description of vector resonances is provided by the
vector or antisymmetric tensor fields. It was shown that
Lagrangians of these two formulations are not equivalent unless
some contact terms are included (it was also proved in \cite{KNT}
that in the general case an infinite number of such terms is necessary).
Another possibility of the description of spin-1 resonances is the so-called
first-order formalism investigated in \cite{KNT}, where
both types of fields are used. For illustrative purposes we
restrict our discussion in the following to the antisymmetric tensor case.

The nonet of vector resonances $1^{--}$ can be represented  by the
antisymmetric tensor fields collected in the $3\times 3$ matrix
$R_{\mu\nu}$

\begin{equation}
  R_{\mu\nu} = \left(
    \begin{array}{ccc}
      \frac{1}{\sqrt2}\rho^0+\frac{1}{\sqrt6}\omega_8+\frac{1}{\sqrt3}\omega_0 & \rho^+ & K^{*+} \\
      \rho^- & -\frac{1}{\sqrt2}\rho^0+\frac{1}{\sqrt6}\omega_8+\frac{1}{\sqrt3}\omega_0 & K^{*0} \\
      K^{*-} & \overline{K}^{*0} & -\frac{2}{\sqrt6}\omega_8+\frac{1}{\sqrt3}\omega_0 \\
    \end{array}
  \right)_{\mu\nu}.
\end{equation}
These fields transform under the nonlinear realization of the
$U(3)_L\times U(3)_R$.

Let us start with  the following Lagrangian
\begin{equation}
  {\cal L} = -\frac12 \langle \partial_\mu R^{\mu\nu}\partial^\rho
  R_{\rho\nu}\rangle + \frac14 M^2\langle
  R_{\mu\nu}R^{\mu\nu}\rangle + {\cal L}_{int},
\end{equation}
where the brackets denote the trace over group indices. In general
the full two-point 1PI Green function has the form\footnote{Here
and in what follows we omit the group indices and trivial group
factors $\delta^{ab}$ for simplicity.}
\begin{equation}
\Gamma _{\mu \nu \alpha \beta }^{(2)}(p)=\frac 12(M^2+\Sigma
^T(p^2))\Pi _{\mu \nu \alpha \beta }^T+\frac 12(M^2-p^2+\Sigma
^L(p^2))\Pi _{\mu \nu \alpha \beta }^L\,, \label{gamma}
\end{equation}
with the projectors
\begin{eqnarray}
\Pi _{\mu \nu \alpha \beta }^T &=&\frac 12\left( P_{\mu \alpha
}^TP_{\nu
\beta }^T-P_{\nu \alpha }^TP_{\mu \beta }^T\right)\,, \\
\Pi _{\mu \nu \alpha \beta }^L &=&\frac 12\left( g_{\mu \alpha
}g_{\nu \beta }-g_{\nu \alpha }g_{\mu \beta }\right) -\Pi _{\mu \nu
\alpha \beta }^T\,,
\end{eqnarray}
where $P_{\mu\nu}^T = g_{\mu\nu} - p_\mu p_\nu/p^2$. Inverting
(\ref{gamma}) we get for the propagator
\[
\Delta _{\mu \nu \alpha \beta }(p)=-\frac 2{p^2-M^2-\Sigma
^L(p^2)}\Pi _{\mu \nu \alpha \beta }^L+\frac 2{M^2+\Sigma
^T(p^2)}\Pi _{\mu \nu \alpha \beta }^T.
\]
This propagator has two types of (generally complex) poles $s_V$
and $s_\VV$. The first one satisfies the equations
\begin{eqnarray}
s_V-M^2-\Sigma ^L(s_V) =0  \label{MV}
\end{eqnarray}
and (assuming $s_V=M_V^2>0$) we have for $p^2\rightarrow M_V^2$
\begin{eqnarray*}
\Delta _{\mu \nu \alpha \beta }(p)
=\frac{Z_V}{p^2-M_V^2}\sum_\lambda u_{\mu \nu }^{(\lambda
)}(p)u_{\alpha \beta }^{(\lambda )}(p)^{*}+O(1)\,,
\end{eqnarray*}
where
\[
Z_V=\frac 1{1-\Sigma ^{^{\prime }L}(M_V^2)}.
\]
The wave function $u_{\mu \nu }^{(\lambda )}(p)$ is expressed in
terms of the spin-one polarization vectors $\varepsilon _\nu
^{(\lambda )}(p)$ as
\[
u_{\mu \nu }^{(\lambda )}(p)=\frac{\mathrm{i}}{M_V}\left( p_\mu
\varepsilon _\nu ^{(\lambda )}(p)-p_\nu \varepsilon _\mu
^{(\lambda )}(p)\right).
\]
Therefore, under the conditions $M_V^2>0$ and $Z_V>0$ such a pole
corresponds to the spin-one state $|p,\lambda,V\rangle$ which
couples to $R_{\mu\nu}$ as
\begin{equation}
\langle 0|R_{\mu \nu }(0)|p,\lambda ,V\rangle =Z_V^{1/2}u_{\mu \nu
}^{(\lambda )}(p).
\end{equation}
One of the solutions of equation (\ref{MV}) is perturbative
$$
  s_V = M^2 + \delta M_V^2\,, \qquad\qquad Z_V = 1 +\delta Z_V\,,
$$
where $\delta M_V^2$ and $\delta Z_V$ are small corrections
vanishing in the free field limit. This solution corresponds to
the original degree of freedom described by the free part of the
Lagrangian  ${\cal{L}}_0$. The other possible non-perturbative solutions of
(\ref{MV}) decouple in the limit of vanishing interaction.

Additional type of poles, given by the solutions of
\begin{equation}
M^2+\Sigma ^T(s_\VV) =0\,, \label{MA}
\end{equation}
is of a non-perturbative nature. For $s_\VV=M_\VV^2>0$ and $p^2\rightarrow
M_\VV^2$ we get
\begin{eqnarray*}
\Delta _{\mu \nu \alpha \beta }(p)
=\frac{Z_\VV}{p^2-M_\VV^2}\sum_\lambda \widetilde{u}_{\mu \nu
}^{(\lambda )}(p)\widetilde{u}_{\alpha \beta }^{(\lambda
)}(p)^{*}+O(1)\,,
\end{eqnarray*}
where
\begin{eqnarray*}
Z_\VV &=&\frac 1{\Sigma ^{^{\prime }T}(M_\VV^2)}\,, \\
\widetilde{u}_{\mu \nu }^{(\lambda )}(p)&=&\frac 12\varepsilon
_{\mu \nu \alpha \beta }u^{(\lambda )\alpha \beta }(p).
\end{eqnarray*}
Assuming therefore  that $M_\VV^2>0$ and $Z_\VV>0$, such a pole
corresponds to the spin-one particle states
$|p,\lambda,\VV\rangle$ (with opposite parity w.r.t.
$|p,\lambda,V\rangle$) which couple to the antisymmetric tensor
field as
\begin{equation}
\langle 0|R_{\mu \nu }(0)|p,\lambda ,\VV\rangle
=Z_\VV^{1/2}\widetilde{u}_{\mu \nu }^{(\lambda )}(p).
\end{equation}
In the free field limit $\Sigma^T(p^2)=0$ and the additional
degrees of freedom are frozen.

The previous discussion suggests that the general form of the
interaction Lagrangian can cause a dynamical generating of
additional degrees of freedom at the one loop level. However, the
general picture is a little bit more subtle. The point is, that
the poles described above  might correspond to negative norm
ghosts (for $Z_V, Z_\VV <0$) or tachyons (for $M_V^2, M_\VV^2<0$)
(see \cite{KNT3} for details).

As a toy example let us assume a simple ``interaction'' Lagrangian
of the form
\begin{equation}
  {\cal L}_{int} = \frac{\alpha}{4}\langle \partial_\alpha
  R^{\mu\nu} \partial^\alpha R_{\mu\nu}\rangle,
  \label{alphaterm}
\end{equation}
which represents actually another type of kinetic term. It
generates contribution to both self-energies $\Sigma^{L,T}(p^2)$
\begin{equation}
  \Sigma^T(p^2) =\Sigma^L(p^2) =  \alpha p^2.
\label{alpha}
\end{equation}
The two solutions of the equations (\ref{MV},\,\ref{MA}) are
therefore the perturbative one
\begin{eqnarray} M_V^2=M^2(1+\alpha+\dots)\,,\qquad Z_V=(1+\alpha+\dots)
\end{eqnarray}
and the non-perturbative one
\begin{eqnarray}
M_\VV^2=-\frac{M^2}{\alpha}\,,\qquad Z_\VV=\frac{1}{\alpha}\,.
\end{eqnarray}
Thus for $\alpha<0$ the additional negative norm ghost is
propagated (tachyon for the case $\alpha>0$). Note that, the
``interaction'' term (\ref{alphaterm}) is not present at the tree
level, however, it can be generated as a counterterm in the
renormalization procedure as we will see in the next section.

\section{One loop contribution}

In order to avoid lengthy expressions, let us concentrate on the
effect of   just one special term of the  interaction
Lagrangian\footnote{ The complete list of terms in even intrinsic
parity sector can be found in \cite{Spain2}, the part of the basis
for odd intrinsic parity sector is provided in \cite{Spain1}.}
with two resonance fields
\begin{equation}
  {\cal L}_{int} = d_1 \epsilon_{\mu\nu\alpha\sigma} \langle D_\beta
  u^\sigma \{R^{\mu\nu},R^{\alpha\beta}\}\rangle+\dots.
  \label{Lint}
\end{equation}
The most general result  will be published in \cite{KNT3} but it
does not differ in essence from what follows.

\begin{figure}
\includegraphics{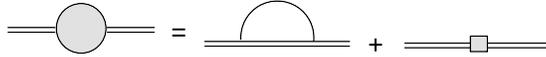}
\caption{\label{fig} The one loop correction to tensor
self-energy. The double line stands for resonance fields, the
single line stands for Goldstone bosons. The two graphs  on the
r.h.s. represent the loop and counterterm contributions
respectively.}
\end{figure}

The explicit calculation (using dimensional
regularization\footnote{In order to avoid the problems with
$d-$dimensional Levi-Civita tensor,  we use the simplest form of
dimensional regularization (known as Dimensional Reduction) by
means of performing first the four-dimensional tensor algebra and
only then regularizing the remaining integrals. The infinite part
of the result does not depend on this choice.}) of the first
Feynman diagram depicted in Fig. 1 with vertices corresponding to
the interaction term (\ref{Lint}) gives for the self-energies
$\Sigma^T(p^2)$ and $\Sigma^L(p^2)$,
\begin{equation}
  \Sigma^T_{loop}(p^2) = \Sigma^L_{loop}(p^2) =
  \frac{5}{6}\, d_1^2\left(\frac{M}{F}\right)^2\frac{d-2}{d}\frac{\mu^{d-4}}{\pi^2}\left(\frac{2}{d-4}+\gamma_E -\ln 4\pi -1+\ln\frac{M^2}{\mu^2}\right)(p^2+M^2)+\dots,
\end{equation}
In order to cancel the UV divergences it is necessary to add to
(\ref{Lint}) the following counterterms
\begin{equation}
  {\cal L}_{ct} = \frac{1}{4}\delta M^2 \langle R_{\mu\nu} R^{\mu\nu} \rangle +
  \frac{\alpha}{4} \langle D^\alpha R_{\mu\nu} D_\alpha R^{\mu\nu}\rangle
  +\frac{\beta}{2}\langle D^\alpha R_{\alpha\mu} D_\beta
  R^{\beta\mu}\rangle+\dots,
\end{equation}
{\em{i.e.}} a mass term and two kinetic terms  one of which was
not present in the original leading order Lagrangian. These
counterterms contribute to $\Sigma^T(p^2)$ and $\Sigma^L(p^2)$ as
(cf. (\ref{gamma}) and (\ref{alpha}))
\begin{eqnarray}
  \Sigma^T_{ct}(p^2) &=&\delta M^2+\alpha p^2\,,\\
  \Sigma^L_{ct}(p^2) &=&\delta M^2+(\alpha+\beta) p^2\,,
\end{eqnarray}
the infinite parts of which are fixed as
\begin{eqnarray*}
  \delta M^2 &=&- \frac{40}{3}\, d_1^2 M^2\left(\frac{M}{F}\right)^2\lambda_\infty + (\delta
  M^2)^r(\mu)+\dots,\\
   \alpha &=&-
  \frac{40}{3}\, d_1^2\left(\frac{M}{F}\right)^2\lambda_\infty +
  \alpha^r(\mu)+\dots,\\
  \beta&=&\beta^r(\mu)+\dots\,,
\end{eqnarray*}
where
$$
  \lambda_\infty = \frac{\mu^{d-4}}{16\pi^2}\left(\frac{1}{d-4}-\frac{1}{2}(-\gamma_E +\ln 4\pi +1)\right).
$$
We see that the  interaction Lagrangian in the antisymmetric
formulation of Resonance Chiral Theory can lead to the nontrivial
momentum dependence of $\Sigma^T(p^2)$, and therefore to the
possible presence of  additional poles which correspond to
opposite parity asymptotic  states or resonances or even negative
norm ghosts or tachyons.

It can be shown that not only the antisymmetric formalism but also
the vector formalism (the additional poles  are spin-0 modes) and the
first-order formalism (where the structure of states is much
richer) suffer from this feature. In \cite{KNT3} the complete
calculation in all three formalisms will be published with
complete Lagrangians up to ${\cal O}(p^6)$.

\section{Conclusion}

In this article we have illustrated the problems connected with
the one-loop renormalization of the propagators of spin-1
resonances within the antisymmetric tensor formulation of
R$\chi$T. As we have shown by means of explicit calculation,  the
renormalization of the theory at one loop level  needs
counterterms including a new type of kinetic term connected with
possible propagation of additional degrees of freedom. In some
cases, these could correspond to negative norm states or tachyons.
Analogous feature can be seen also in alternative formulations of
R$\chi$T with spin-1 resonances described by vector fields or by
the first-order formalism \cite{KNT3} and can be understood as a
manifestation of the well-known fact that, without gauge symmetry
and Higgs mechanism, the quantum field theory of massive spin-1
particles might suffer from internal inconsistencies (for a recent
discussion see {\em e.g.} \cite{gH} and references therein).

In all cases, in order to vindicate R$\chi$T as a useful effective
quantum field theory, we have to take into account  this
phenomenon. More detailed discussion will be published in
\cite{KNT3}.

\begin{acknowledgments}
We are grateful to J.~J.~Sanz-Cillero and R.~Rosenfelder for valuable
discussions and comments to the text.
This work was supported in part by the Center for Particle Physics
(project no. LC 527), GACR (grant no. 202/07/P249) and by the EU
Contract No. MRTN-CT-2006-035482, \lq\lq FLAVIAnet''.
\end{acknowledgments}


\end{document}